\documentclass[doublecol]{epl2} 

\usepackage{graphicx,color}

\usepackage{amsmath}
\usepackage{hyperref}
\usepackage{cite}
\usepackage{xcolor}
\hypersetup{colorlinks,bookmarksopen,bookmarksnumbered,
citecolor=red,
linkcolor=blue,
pdfstartview=green,
urlcolor=teal}

\newcommand{\be}{\begin{equation}}
\newcommand{\ee}{\end{equation}}
\newcommand{\bea}{\begin{eqnarray}}
\newcommand{\eea}{\end{eqnarray}}

\newcommand{\e}{\epsilon}

\title{Full counting statistics in the gapped XXZ spin chain}

\author{Pasquale Calabrese\inst{1,2,3} \and Mario Collura\inst{1} \and Giuseppe Di Giulio\inst{1,2} \and Sara Murciano\inst{1,2}}
\shortauthor{P. Calabrese, M. Collura, G. Di Giulio, S. Murciano}

\institute{                    
  \inst{1} International School for Advanced Studies (SISSA), via Bonomea 265, 34136  Trieste, Italy\\
  \inst{2} INFN, sezione di Trieste, via Bonomea 265, 34136  Trieste, Italy\\
  \inst{3} International Centre for Theoretical Physics (ICTP), I-34151, Trieste, Italy 
}

\abstract{
We exploit the knowledge of the entanglement spectrum in the ground state of the gapped XXZ spin chain to derive asymptotic exact results for the full counting statistics 
of the transverse magnetisation in a large spin block of length $\ell$. 
We found that for a subsystem of even length the full counting statistics is Gaussian, while for odd subsystems it is the sum of {\it two} Gaussian distributions. 
We test our analytic predictions with accurate tensor networks simulations. 
As a byproduct, we also obtain the symmetry (magnetisation) resolved entanglement entropies. 
}

\begin{document}

\maketitle

\section{Introduction}
The process of measurement in quantum mechanics is intrinsically probabilistic:  the measure of a given observable generically 
provides different outcomes in identically prepared systems.
Hence, the probability distribution (PDF) of an observable is a natural quantity to consider in any quantum mechanical system and provides
much more information than the average value of the same observable. 
In many-body systems, these PDFs, or equivalently their full counting statistics (FCS), have been the subject of intensive investigations since many years with a focus mainly 
on local observables (i.e. defined in a given point or lattice site) or global ones (i.e. extensive quantities involving the entire system). 
Only in recent time, the attention shifted to observables with support on a finite, but large, subsystem embedded in a thermodynamic one, partially motivated by 
some cold atomic experiments \cite{HLSI08,KPIS10,KISD11,GKLK12,AJKB10,JABK11} and by the connection with the entanglement entropy of the same subsystem  
\cite{kl-09a,kl-09b,hgf-09,sfr-11a,sfr-11b,cmv-12,lbb-12,si-13,clm-15,u-19}. 
In spite of a large recent literature on the subject 
\cite{cd-07,bss-07,aem-08,lp-08,ia-13,sk-13,e-13,k-14,mcsc-15,sp-17,CoEG17,nr-17,hb-17,bpc-18,gadp-06,er-13,lddz-15,cdcd-20,c-19,bp-18,ce-20,ag-19,gec-18,nr-20,ppg-19}, 
results based on integrability for one-dimensional exactly solvable {\it interacting} models  are still scarse (see \cite{bpc-18,bp-18}).

\begin{figure*}[t]
\begin{center}
\includegraphics[width=\textwidth]{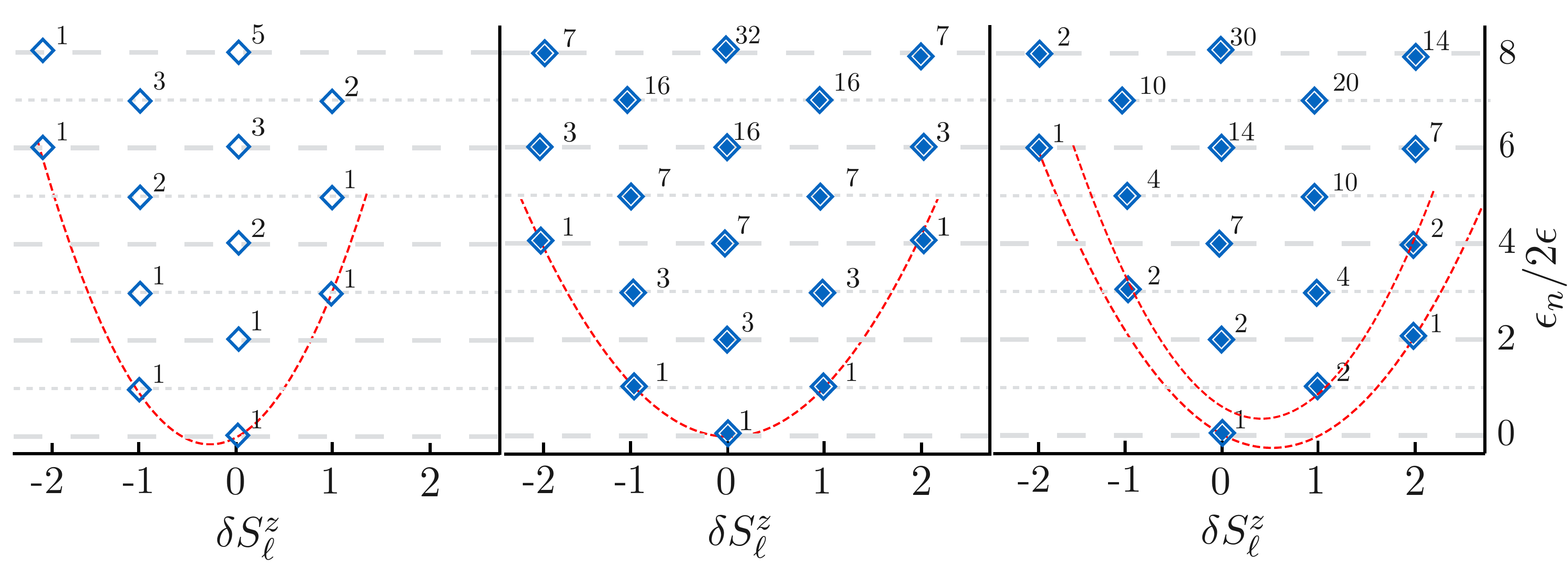}
\end{center}
\caption{%
Entanglement spectra of the gapped XXZ spin chain in the three configurations we consider here. 
Left: Semi-infinite line. Center: A block of $\ell$ contiguous spins with $\ell$ even. Right: A block with $\ell$ odd. 
We report the logarithm of the eigenvalues of the reduced density matrix $\e_n$ in units of $2\e$, with $\e={\rm arccosh}\Delta$, as
function  of $\delta S_\ell^z$ (cf. \eqref{deltaS}). 
Each tilted square signals the presence of an eigenvalue with degeneracy given by the nearby number.  
The dashed-red parabolas are envelopes of the location of the largest eigenvalue of the RDM at fixed  $\delta S_\ell^z$. 
Notice that in the left and in the center, the towers of degeneracies are independent of $\delta S_\ell^z$.
Conversely, on the right, i.e. for odd blocks, there are two towers depending on the parity of $\delta S_\ell^z$. 
}
\label{Fig:spec}
\end{figure*}

In this Letter, we provide an explicit exact calculation for the PDF and for the FCS of an observable within an extended subsystem. 
We consider the ground state of the XXZ spin chain defined by the Hamiltonian
\begin{equation}
\label{eq:XXZ Hamiltonian}
H_{\mathrm{XXZ}}
=
\sum_{j}
\left[
\sigma_j^x \sigma_{j+1}^x
+
\sigma_j^y \sigma_{j+1}^y
+
\Delta
\sigma_j^z \sigma_{j+1}^z
\right],
\end{equation}
where $\sigma^\alpha_j$, $\alpha=x,y,z$ are the Pauli matrices at site $j$. 
%
We focus in the antiferromagnetic gapped regime with $\Delta>1$.
The observable of interest is the transverse magnetisation for a block of $\ell$ contiguous spins, i.e. 
\be
S_\ell^z=\frac12\sum_{j=1}^{\ell} \sigma^z_j.
\ee
In particular, since the total transverse magnetisation ($\sum_j \sigma_j^z/2$) is conserved, the FCS can be directly obtained from the entanglement spectrum of the subsystem.
Indeed, the reduced density matrix $\rho_\ell$ of the subsystem is organised in blocks of fixed magnetisation (quantised in terms of integers or half-integers up to $\ell/2$
depending on the parity of $\ell$). 
In order to work with an observable with integer eigenvalues for any $\ell$, 
it is convenient to focus on the difference of the block magnetisation with the N\'eel state, i.e.
\be
\delta S^z_\ell\equiv \sum_{j=1}^{\ell} \Big (\frac{\sigma^z_{j}}2 - \frac{(-1)^j}2\Big)\,.
\label{deltaS}
\ee 
The probability of a measurement of the subsystem magnetisation with outcome $\delta S^z_\ell=q$ is just the trace of the block of $\rho_\ell$ in the sector with $\delta S^z=q$, i.e.
\be
P(q)= {\rm Tr} \rho_\ell \Pi_q= \sum_{s\,\in\, \mathcal{S}_q} \lambda_s\,,
\ee
where $\Pi_q$ is the projector on the sector of magnetisation $\delta S^z=q$, $\lambda_s$ are the eigenvalues of $\rho_\ell$,  and $\mathcal{S}_q$ stands for 
all the eigenvalues in that magnetisation  sector  (notice that $\sum_q P(q)={\rm Tr} \rho_\ell=1$ by construction).
Similarly the FCS generating function is defined as 
\begin{equation}
\label{eq:FourierSeriesDef}
G(\lambda)\equiv {\rm Tr} \rho_\ell e^{i \lambda \delta S_\ell^z}= 
\sum_q P(q) e^{i q \lambda};
\end{equation}
its derivatives in $\lambda=0$ provide the moments of the observables $\delta S_\ell^z$. 
Hence the exact knowledge of the entanglement spectrum also provides the FCS of the total transverse magnetisation (in general it provides the FCS of any
conserved charge).
For the ground state of the XXZ spin chain in the gapped regime, the entanglement spectrum has been obtained in Ref. \cite{albaES}. 
We exploit its knowledge here to reconstruct the PDF and the FCS of the subsystem magnetisation.

The remaining of this Letter is organised as follows. 
First, we recap and generalise results for the entanglement spectrum of Ref. \cite{albaES}.
Then we reconstruct the PDF and the FCS for both even and odd number of sites of the subsystem.
As a byproduct, we also derive some results for the symmetry resolved entanglement. 
Finally we draw our conclusions.

\section{Recap on the entanglement spectrum} 
We consider the symmetry broken ground state, i.e. the one that for large $\Delta$ converges to the N\'eel state.
This state is doubly degenerate, so there are two equivalent states which are mapped into each other by the translation of one site. 
Let us first consider the case of $\ell=\infty$, i.e. the subsystem being the semi-infinite line. 
In this case, the logarithm of the eigenvalues of $\rho_\ell$ are equispaced, i.e. $\lambda_s=e^{-\epsilon_s}$ 
with $\epsilon_s = 2\epsilon s$, where $\epsilon={\rm arccosh}\Delta$ (see, e.g., \cite{peschel1,ccp-10}).
The total degeneracy $D_{\rm h}(s)$ of the level $2\epsilon s$ is 
the number of partitions of $s$ into smaller non-repeated integers (including zero). 
Here we need to know how these eigenvalues distribute among sectors of fixed magnetisation; 
this has been worked out in \cite{albaES} with a combination of perturbation theory and integrability arguments. 
The first panel of Figure \ref{Fig:spec} reports the structure of the entanglement spectrum based on the results of Ref.  \cite{albaES}. 
%
The final result for the degeneracy of the eigenvalue with $\delta S^z_\ell=q$ at level $s$
is $d_{\rm h}(q,s)= p_{\rm h}(\frac{s-m_{\rm h}(q)}{2})$  \cite{albaES}, with $p_{\rm h}(n)$ the number of integer partitions of 
$n$ and $m_{\rm h}(q)=q(2q+1)$. 
We use the convention that $p_{\rm h}(x)=0$ for negative integers and for half-integers. 
(The other degenerate state --sometimes called Antineel for $\Delta\to\infty$-- is obtained by sending $q\to-q$ with the net effect of having $m_{\rm a}(q)=q(2q-1)$.)
The number of partitions $p_{\rm h}(x)$ has not an analytic form; the same is true for the total degeneracy $D_{\rm h}(x)$ above; 
however both have simple generating functions given by
\begin{equation}
\label{eq:generatingfcts}
\sum_{s=0}  p_{\rm h}(s) x^s=
\prod_{k=1}^\infty \frac{1}{1-x^k},
\qquad
\sum_{s=0}  D_{\rm h}(s) y^s =
\prod_{k=1}^\infty (1+y^k).
\end{equation}
Notice that the degeneracies of the various sectors are all the same, there is only an overall shift of the lowest eigenvalues at fixed $q$ given by $m_{\rm h}(q)$.
Recalling that the PDF $P(q)$  is just the sum of all the eigenvalues of the RDM at fixed $q$ (weighted with their degeneracy),
this is the same for all $q$ except for the important factor of the largest eigenvalue at fixed $q$ equal to $e^{-2\e m_{\rm h}(q)}$.   
Hence the PDF is just
\be
\label{PDF1}
P_{\rm h}(q)= {\cal N}  e^{-2 \epsilon q(2q+1)}\,,  \quad {\rm with} \;
{\cal N}^{-1}= 
\theta_3\left(i\e ,e^{-{4\e}}\right),
\ee
and indeed it was already obtained in Ref. \cite{mdc-20}  (here $\theta_3$ is the elliptic theta function).

Now, still following the approach of Ref. \cite{albaES}, we explain how to use these results to obtain the entanglement spectrum of a finite large interval. 
%
As long as $\ell$  is larger than the correlation length, 
the reduced density matrix $\rho_\ell$ of a single interval with two boundaries factorises into $\rho_L\otimes \rho_R$, where $\rho_{L/R}$ are the reduced density matrices 
for the semi-infinite lines having the left/right end-point of the interval. 
The combination of these two spectra into a single one is graphically reported in Fig. \ref{Fig:spec}.
We show both cases for even and odd subsystems (center and right respectively). 
For an even subsystem we should combine two different spectra $m_L(x) = m_{\rm h}(x)= x(2x+1)$ and $m_R(x) =m_{\rm a}(x)= x(2x-1)$ from left and right.
Conversely, for odd blocks, the left and the right spectra to combine are equal, e.g.  $m_L(x) = m_R(x) = x(2x+1)$.
The final results for the degeneracies are reported in the figure. 
In the even case, we have that the degeneracy at fixed $q$ at level $s$ can be written as $d_{\rm e}(q,s)= p_{\rm e} (\frac{s-m_{\rm e}(q)}{2})$ with 
 $m_{\rm e}(q)=q^2$ and $p_{\rm e}$ generated by
\be
\sum_{s=0}   p_{\rm e}(s) x^s=\prod_{k\geq1} \frac{(1 + x^k)^2}{1 - x^k}\,,
\label{gen_even}
\ee
leading to the generating function for the total degeneracy $D_{\rm e}(s)$ of level $s$
\be
\sum_{s=0} D_{\rm e}(s) x^s=\prod_{k\geq1}  (1+x^k)^2 .
\label{gen_even_tot}
\ee
Notice that while the generating function for $D_{\rm e}(s)$ is the square of the one for $D_{\rm h}(s)$, the same is not true for $p_{\rm e}$. 
Again we employ the convention $p_{\rm e}(x)=0$ for negative numbers and for half-integers.

For odd blocks, it is more complicated to combine the two spectra for even and odd $q$. The degeneracies of both sectors have the generating function
\be
\sum_{s=0}   p_{\rm o}^{\rm b}(s) x^s=\prod_{k\geq1}  \frac{(1-x^{2 k})^3}{(1-x^k)^2 (1-x^{4 k})^2},
\label{gen_odd}
\ee
where even (odd) powers of $x$ correspond to even (odd) values of $q$.
However, a single generating function for different $q$ is not a too useful tool to write symmetry resolved quantities. 
Exploiting some identities of elliptic theta functions $\theta_{2,3}$,  we can extract the even and the odd part of \eqref{gen_odd} we are interested in. 
After some algebra we get  (for $x>0$)
\be
\sum_{s=0}   p_{\rm o}(s,q) x^s=
\frac{(\theta _2(x^4))^{\frac{1-(-1)^q}{2}} (\theta_3(x^4))^{\frac{1+(-1)^q}{2}}}{\prod_{k\geq1}(1+x^{2 k}) (1-(-x)^k) (1-x^k) },
\label{gen_odd2}
\ee
which does depend on the parity of $q$. 
Hence the degeneracy of the level $s$ with fixed $q$ is
\be
d_{\rm o}(q,s)= p_{\rm o}(s-m_{\rm o}(q)),
\ee
with $m_{\rm o}(q)=q^2-q$. Indeed $m_{\rm o}(q)$ and $m_{\rm o}(q)+1$ are the two parabolas in Fig. \ref{Fig:spec}, envelopes of the largest eigenvalues of 
the RDM for even and odd $q$ respectively.
The generating function for the total degeneracy $D_{\rm o}(s)$ of level $s$ is the same as $D_{\rm e/h}(s)$ in Eq. \eqref{gen_even_tot}.

\begin{figure*}[t]
\begin{center}
\includegraphics[width=0.47\textwidth]{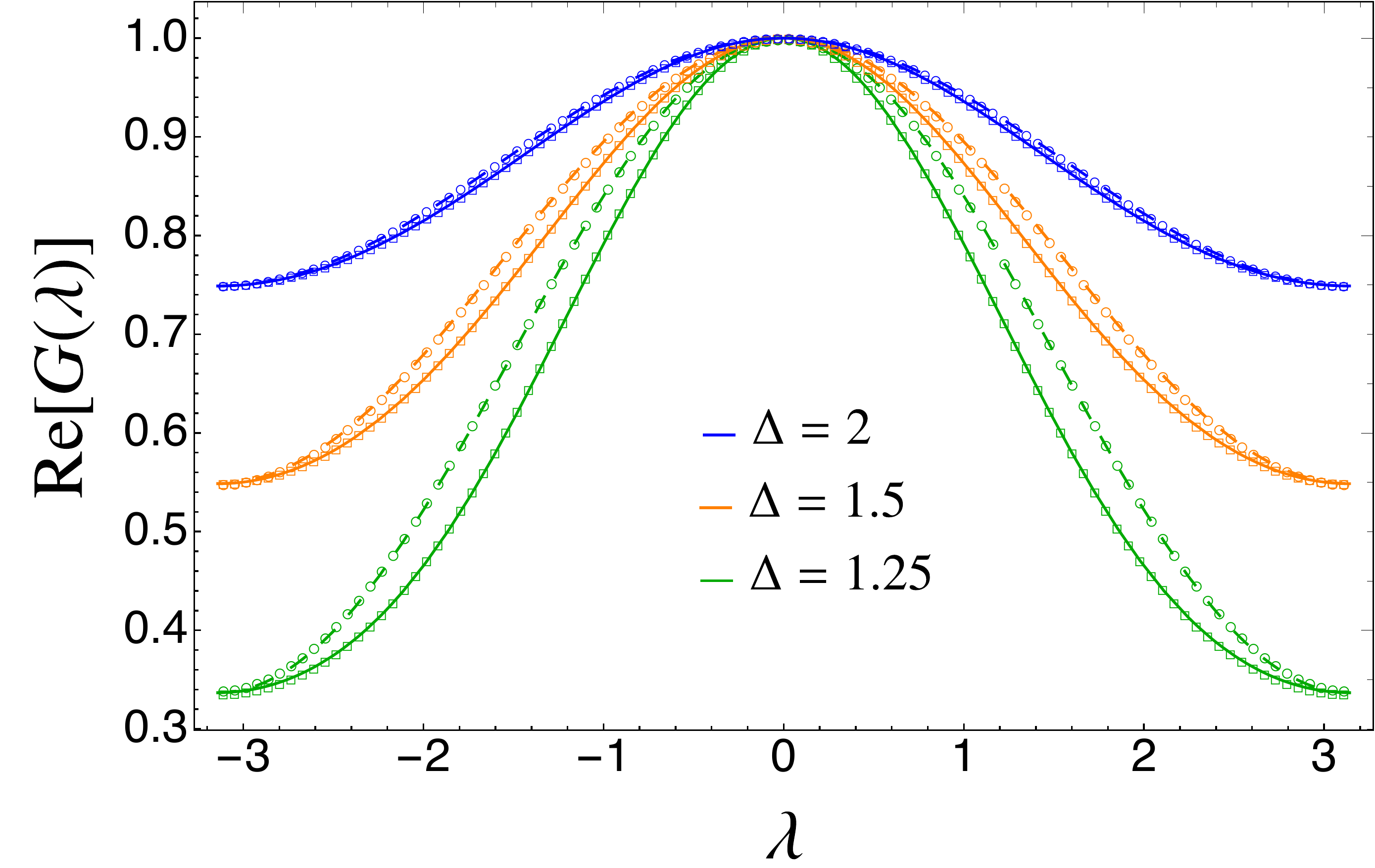}
\includegraphics[width=0.47\textwidth]{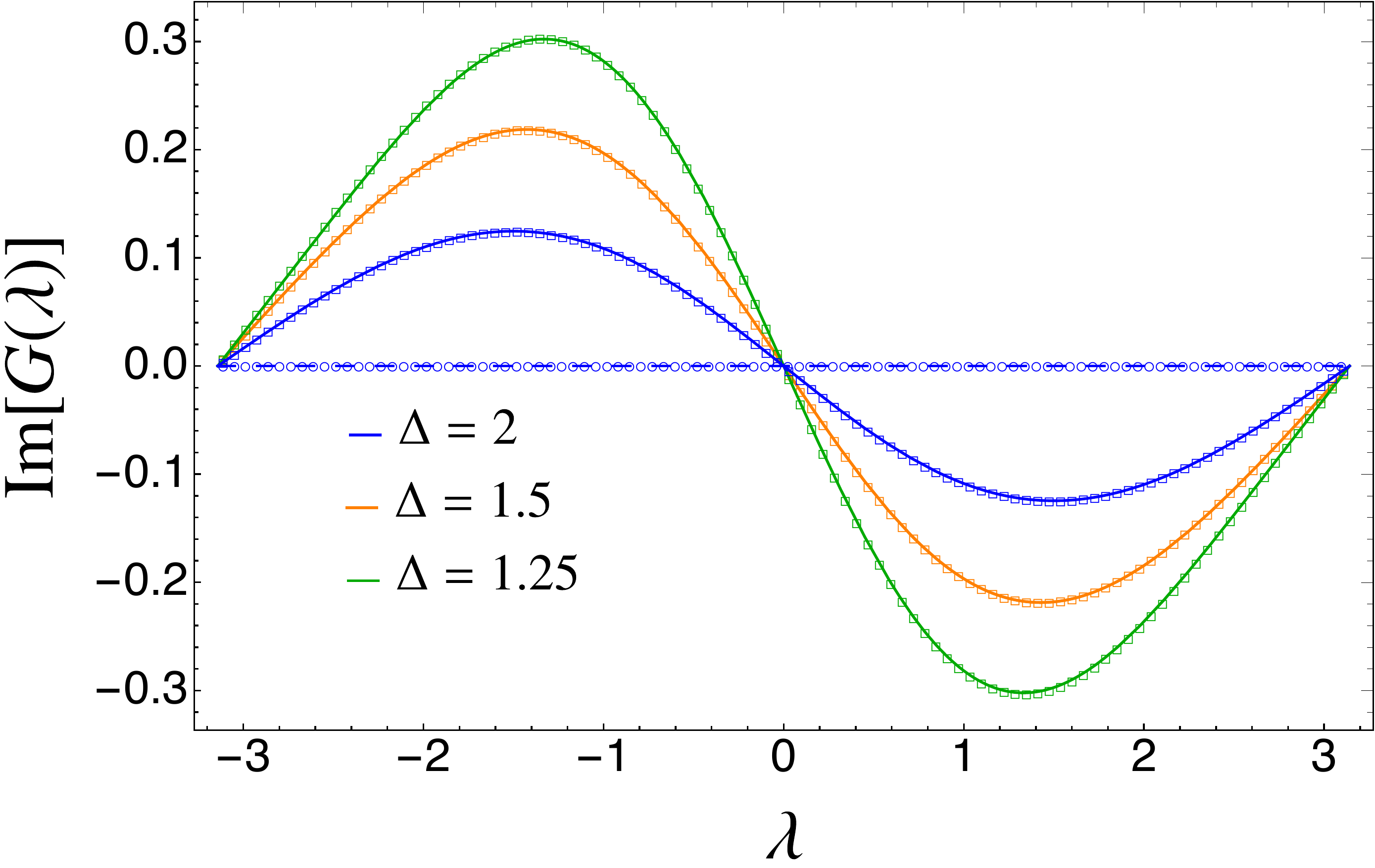}
\end{center}
\caption{%
Full counting statistics generating functions $G(\lambda)$ for the gapped XXZ spin chain for three values of $\Delta$.
The left (right) panel is the real (imaginary) part of  $G(\lambda)$.
The symbols are the iTEBD data that perfectly match the superimposed analytic  predictions (full lines for odd $\ell$ and dashed for even $\ell$).
The data are for infinite chains and subsystems equal to $\ell =200$ (circles) or $201$ (squares).  
Notice that the real parts for even and odd $q$ are qualitatively similar, but quantitatively different. 
}
\label{Fig:FCS}
\end{figure*}

\section{Full counting statistics: even number of sites}
The easiest way to get the PDF $P_{\rm e}(q)$ for the interval  is to combine the PDFs at the right and left boundary as
\be
P_{\rm e}=\sum_{q_1=-\infty}^\infty P_L(q_1)P_R(q-q_1)=  \sum_{q_1=-\infty}^\infty P_{\rm h}(q_1)P_{\rm h}(q_1-q), 
\ee
where we used that the PDF at the two boundaries are $P_L(q)= P_{\rm h}(q_1)$ and $P_R(q)=P_{\rm h}(-q)$. 
The sum is easily rewritten as 
\be
P_{\rm e}(q)= {\cal N}^2 e^{-2\e (q^2-1/4)} \sum_{q_1=-\infty}^\infty e^{-2\e (2q_1+ q+1/2)^2}\,.
\ee
The remaining sum over $q_1$ does not depend on $q$, for integer $q$. 
Hence the PDF is Gaussian
\be
P_{\rm e}(q)={\cal N}_{\rm e} e^{-2q^2 \epsilon},
\ee
and the normalisation factor is $ {\cal N}_{\rm e}^{-1}=\sum_{q} e^{-2q^2 \epsilon}= \theta_3(e^{-{2 \e}})$. %
The FCS is  the Fourier series \eqref{eq:FourierSeriesDef} which immediately leads to
\be
G_{\rm e} (\lambda)= \frac{\theta_3\left(\frac{ \lambda}{2},e^{-{2 \e}}\right)}{\theta_3(e^{-{2 \e}})}.
\label{Ge}
\ee
Notice that this is real and even in $\lambda$. 
As a cross check, the same result is re-obtained by directly summing over the eigenvalues of the RDM 
with the degeneracies reported in Fig. \ref{Fig:spec} (to perform the sum, one exploits \eqref{gen_even}  the product representation of the
$\theta_3$ function).

The FCS generating function is directly measured in iTEBD simulations \cite{TEBD}, as explained  in details, e.g., in Refs. \cite{CoEG17,ce-20}. 
The results in the thermodynamic limit for three values of $\Delta>1$ and for $\ell=200$ are shown in Figure \ref{Fig:FCS}. 
The agreement is always excellent (data and predictions are superimposed) for all considered values of $\Delta$.
We mention that as $\Delta$ gets close to $1$, one should consider much larger values of $\ell$ to reach such good agreement due to the diverging correlation length
at the isotropic point. 

\section{Full counting statistics: odd number of sites}
Also for  this case, the PDF can be obtained combining two single-boundary ones as
\be
P_{\rm o}=\sum_{q_1=-\infty}^\infty P_L(q_1)P_R(q-q_1)=  \sum_{q_1=-\infty}^\infty P_{\rm h}(q_1)P_{\rm h}(q-q_1), 
\ee
where we used that the PDF at the two boundaries are the same. 
Again, the sum is easily rewritten as 
\be
P_{\rm o}(q)= {\cal N}^2 e^{-2\e (q^2-q)} \sum_{q_1=-\infty}^\infty e^{-2\e (2q_1-q)^2}\,.
\ee
However, this time the remaining sum {\it does} depend on the parity of $q$. Performing this sum, the PDF is 
\be
P_{\rm o}(q)= {\cal N}_{\rm o} e^{-2\e (q^2-q)}\times
\begin{cases} 
\theta_3(e^{-8\e})\,,  & q\; {\rm even},\\
\theta_2(e^{-8\e})\,,&  q\; {\rm odd}, \\
\end{cases} 
\ee
with ${\cal N}_{\rm 0}$ easily obtained from the normalisation. 

The FCS is  the Fourier series \eqref{eq:FourierSeriesDef} which, after some manipulations using the properties of elliptic functions, leads to
\be
G_{\rm o} (\lambda)= \left(
\frac{\theta_3\big(i \e-\frac{\lambda}2 , e^{-4\e}\big)}{\theta_3(i \e, e^{-4\e})}
\right)^2.
\label{Go}
\ee
Notice that this FCS has a non-vanishing and non-trivial imaginary part, but satisfy $G_{\rm o} (\lambda)^*=G_{\rm o} (-\lambda)$.
Again, as a cross check, this result is re-obtained by directly summing over the eigenvalues of the RDM  with the degeneracies reported in Fig. \ref{Fig:spec}.

Also for odd $\ell$, the analytical prediction \eqref{Go} is tested against iTEBD simulations in Figure  \ref{Fig:FCS}.
In these simulations, we measure the FCS of the operator $S^z_\ell$ and not $\delta S^z_\ell$; 
hence the numerical data have been divided by $e^{i\lambda/2}$. 
After this normalisation, the agreement between data and prediction is extremely good in all considered cases.


\section{Byproduct: symmetry resolved entropies}
A very recent research line in many body quantum systems is to understand how the entanglement organises into the various symmetry sectors of 
a theory \cite{lr-14,fis,goldstein,goldstein1,xavier,goldstein2,riccarda,tr-19,fg-19,clss-19,mdc-20}. 
The reduced density matrix is symmetry decomposed as $\rho_\ell=\oplus_q P(q)\rho_{\ell}(q)$. 
The symmetry resolution of the entanglement spectrum reported in Figure \ref{Fig:spec} allows us to access the symmetry resolved moments as
\be
{\cal Z}_n(q)\equiv \sum_{s\in {\cal S}_q} \lambda_s^n=
\dfrac{\sum_j  d(q,j) e^{- 2 n \epsilon j}}{\big(\sum_{j}  D(j)  e^{- 2  \epsilon j}\big)^n}\,,
\ee
where $d(q,j)$ and $D(j)$ are respectively the degeneracies  of the $j$-th eigenvalue for fixed $q$ and total (whose generating functions are known in the 
three cases of interest). 
The symmetry resolved entropies are defined as
\begin{equation}
\label{eq:RSREE}
S_{n}(q) \equiv \dfrac{1}{1-n}\ln \mathrm{Tr} \rho^n_\ell(q)=  \dfrac{1}{1-n}\ln \mathrm{Tr} \frac{{\cal Z}_n(q)}{{\cal Z}_1^n(q)}.
\end{equation}
Since in $S_n(q)$ only the ratio ${{\cal Z}_n(q)}/{{\cal Z}_1^n(q)}$ matters, the dependence on $D_j$ cancels and 
\be
\frac{{\cal Z}_n(q)}{{\cal Z}_1^n(q)}= \dfrac{\sum_{j}  d(q,j) e^{- 2 n \epsilon j}}{\Big(\sum_{j} d(q,j) e^{- 2  \epsilon j}\Big)^n}
=  \dfrac{\sum_{j}   p(j) e^{- 2 a n \epsilon j}}{\Big(\sum_{j} p(j) e^{- 2a  \epsilon j}\Big)^n}
, \label{ratio}
\ee
where in the last equality we used  $d(q,s)= p(\frac{s-m(q)}{a})$ (with $a=2$ for semi-infinite and even $\ell$, while $a=1$ for odd $\ell$) and 
shifted the sum as $(j-m(q))/a \to j$ (notice that the actual value of $m(q)$ is unessential).

The result for the semi-infinite line has been already derived in Ref. \cite{mdc-20} and we recall it here:
\be
S_n^{\rm h}(q)= \frac1{1-n}{\displaystyle \sum_{k=1}^\infty [ n  \ln (1- e^{-4\e k})-\ln (1- e^{-4n\e k})] } ,
\label{Snh}
\ee
as simply follows combining \eqref{ratio} with \eqref{eq:generatingfcts}. 
Now we derive the entropies  for a finite interval of both even and odd length. 
For even $\ell$, the two sums in (\ref{ratio}) can be rewritten in terms of generating functions \eqref{gen_even}
(with $x= e^{- 4 n \epsilon}$), obtaining 
\begin{equation}
S_n^{\rm e}(q) = \frac{\displaystyle \sum_{k=1}^\infty \Big[\ln \frac{(1+ e^{-4n\e k})^2}{1- e^{-4n\e k}} -n  \ln \frac{(1+ e^{-4\e k})^2}{1- e^{-4\e k}}\Big]}{1-n} .
\end{equation}
Very importantly, the symmetry resolved entropies are {\it not the double} of the single resolved entropies  \eqref{Snh} for the half line
as it is the case for the total one (mathematically this is a consequence of the relation between the generating function for 
 $D_{\rm e}(s)$ and $D_{\rm h}(s)$, but not for $p_{\rm e/h}$). 
 Also, these symmetry resolved entanglement entropies are independent of $q$ and hence satisfy the equipartition of entanglement \cite{xavier} exactly.

In the very same fashion, we can repeat the calculation for odd $\ell$, obtaining the more cumbersome expression
\begin{multline}
 S_n^{\rm o}(q)= \frac1{1-n}
\Bigg[ \\
\sum_{k=1}^\infty  \Bigg(n \ln (1+ e^{-4\e k})(1- e^{-2\e k}) (1-(-)^k e^{-2\e k})  \\ 
-\ln (1+ e^{-4\e nk})(1- e^{-2\e nk}) (1-(-)^k e^{-2\e n k}) \Bigg)\\
+\frac{1+(-)^q}2 ( \ln \theta_3(e^{-8 \e n})- n\ln \theta_3(e^{-8 \e })\\
+
\frac{1-(-)^q}2 ( \ln \theta_2(e^{-8 \e n})- n\ln \theta_2(e^{-8 \e })\Bigg]
,
\end{multline}
Hence, for odd $\ell$, the symmetry resolved entropies do depend on the parity of $q$ and the equipartition of entanglement is explicitly broken.

We finally mention, as a highly non-trivial crosscheck, that it is possible, but cumbersome, to sum over the various sectors $q$ in order to recover the total entanglement, both 
for even and odd $\ell$. The calculation parallels the one for the half-line in \cite{mdc-20}.

\section{Conclusions}
We computed the FCS of the transverse magnetisation in gapped XXZ chains within a spin block of length $\ell$, for $\ell$  larger than the correlation length. 
Our main results are the exact formulas for the generating functions  \eqref{Ge} and \eqref{Go} valid for even and odd $\ell$ respectively.
Their accuracy has been tested against  iTEBD simulations, see Figure \ref{Fig:FCS}. 
The astonishing simplicity of the final results resides in the entanglement spectrum being equispaced, as reported in Figure \ref{Fig:spec}.
The symmetry resolved entanglement entropies turn out to be a simple byproduct of our results.

An extremely interesting open question is whether one can access the crossover from the conformal regime \cite{bss-07,aem-08} (valid for $\ell\ll \xi$, with $\xi$ being the 
correlation length) to massive one ($\ell\gg \xi$) we obtained here.  It is likely that exact techniques developed for the entanglement entropy \cite{ccd-08,cd-09}
may be used even for this problem.


\acknowledgments 
We thank Fabian Essler for  discussions. 
PC and SM acknowledge support from ERC under Consolidator grant  number 771536 (NEMO).

\end{document}